\documentclass[onecolumn,showpacs,nobibnotes,nofootinbib]{revtex4}
\usepackage{amsmath,amssymb,bm}
\usepackage{epsfig}
\usepackage{graphicx}
\usepackage{slashed}
\usepackage{psfig}

\def\lsim{\raise0.3ex\hbox{$<$\kern-0.75em\raise-1.1ex\hbox{$\sim$}}}
\def\gsim{\raise0.3ex\hbox{$>$\kern-0.75em\raise-1.1ex\hbox{$\sim$}}}

\newcommand{\avg}[1]{\left\langle #1 \right\rangle}
\newcommand{\be}{\begin{equation}}
\newcommand{\ee}{\end{equation}}

\def\beq{\begin{equation}}
\def\eeq{\end{equation}}
\def\beqa{\begin{eqnarray}}
\def\eeqa{\end{eqnarray}}

\newcommand{\bb}{\bm{b}}
\newcommand{\rr}{\bm{r}}
\newcommand{\xx}{\bm{x}}
\newcommand{\yy}{\bm{y}}
\newcommand{\zz}{\bm{z}}
\newcommand{\calN}{\mathcal{N}}
\def\gappeq{\mathrel{\rlap {\raise.5ex\hbox{$>$}}
{\lower.5ex\hbox{$\sim$}}}}

\def\lappeq{\mathrel{\rlap{\raise.5ex\hbox{$<$}}
{\lower.5ex\hbox{$\sim$}}}}

\def\Toprel#1\over#2{\mathrel{\mathop{#2}\limits^{#1}}}

\begin{document}

\title{Effects of gluon number fluctuations on
$\gamma\gamma$ collisions at high energies} 
\author{ V.P. Gon\c{c}alves and J. T. de Santana Amaral}
\affiliation{
Instituto de F\'{\i}sica e Matem\'atica,  Universidade
Federal de Pelotas, 
Caixa Postal 354, CEP 96010-900, Pelotas, RS, Brazil}
\begin{abstract}
We investigate the effects of gluon number fluctuations on the total $\gamma\gamma$, $\gamma^*\gamma^*$ cross sections and the  photon structure function $F_2^\gamma(x,Q^2)$. Considering a model which relates the dipole-dipole and dipole-hadron scattering amplitudes, we estimate these observables by using event-by-event and physical amplitudes. We demonstrate that both analyses are able to describe the LEP data, but predict different
behaviours for the observables at high energies, with the gluon fluctuations effects decreasing the cross sections. We conclude that the study of   $\gamma \gamma$ interactions can be useful to constrain the QCD dynamics.
\end{abstract}

\pacs{12.38.-t, 24.85.+p, 25.30.-c}

\maketitle

\section{Introduction}

The high energy evolution of dipole scattering amplitudes in QCD is described
by the pomeron loop equations \cite{IT04,MSW05,IT05,LL05}, a generalization of the Balitsky-JIMWLK hierarchy
(see \cite{GIJV10} and references therein) by including the gluon number fluctuations. In the case when the strong coupling
constant $\alpha_s$ is fixed, the fluctuations wash out the BFKL approximation
and the geometric scaling behavior predicted by Balitsky-Kovchegov (BK)
equation \cite{Balitsky,K99a,K99b}--the simplest (mean-field) nonlinear evolution equation which describes
the evolution of the amplitude for the scattering between a quark-antiquark
pair projectile (a dipole) and a dense target.

In the last decade, many efforts have been made in the investigation of the
fluctuation effects. On the theoretical side,
because of the complexity of the pomeron
loop equations, whose properties up to now have been obtained only under some
approximations, in the last few years fluctuations in high energy evolution
have been studied through simple (toy) models inspired in QCD \cite{Stasto05,KL5,SX05,KozLev06,BIT06,onedim,MS08,MSS08}.
One of such models, which allows the inclusion of both fluctuation and running coupling effects simultaneously, has shown that fluctuations are strongly suppressed by the running of the coupling, up to extremely high energies \cite{DIPS07}. However, it should be pointed out that this result can be model dependent and that the investigation of both effects in real QCD is still a challenge. On the phenomenological side, and in the fixed coupling case, fluctuation effects have been investigated at HERA and RHIC/LHC energies. They have been included in the description of the data on inclusive and
diffractive electron--proton deep inelastic scattering (DIS)
 \cite{Kozlov07,AGBSflu,Xiang09,Xiang10-ddis,F11}. { Although the
results have shown some improvement in the description of the observables, they have not been conclusive with respect to the presence of the fluctuations
in the experimental data}. They have also been studied in the analysis of the pseudo-rapidity distribution of hadron multiplicities of high energy Au+Au collisions at RHIC and in predictions for these observables in Pb+Pb collisions by using Color Glass Condensate dynamics at LHC/ALICE \cite{Xiang10}. It has been found that the charged hadron multiplicities at central rapidity are significantly smaller than saturation based calculations and are compatible to those obtained on a study of multiplicities in the fragmentation region
with running coupling corrections \cite{Albacete07}.

One can see that, up to now, high energy QCD phenomenology in the presence of fluctuations has been studied in few papers only and, particularly, in processes where two scales are present. As is well known,  for $ep/pp$ colliders, the study of the QCD Pomeron is made difficult by the fact that the cross section is
influenced by both short and long distance physics. Only when specific conditions are satisfied is that one can expect
to determine the QCD pomeron effects. Some examples are the forward jet production in deeply inelastic events at
low values of the Bjorken variable $x$ in lepton-hadron scattering and jet production at large rapidity separations in
hadron-hadron collisions, which are characterized by one hard scale. 
This motivates us to look for their effects on different processes.
{ One of these processes} is the off-shell photon scattering at high energy in $e^+\,e^-$ colliders, where the photons are produced from the lepton beams by bremsstrahlung (For a review see, e.g., Ref. \cite{Nisius99}). In these two-photon reactions, the photon virtualities can be made large enough to ensure the applicability of  perturbative methods. Moreover, the photon virtualities can be varied to test the transition between the soft and hard regimes of the QCD dynamics and it is possible to scan the kinematical region to determine the range  where the contribution of the  gluon fluctuations effects is larger. Up
to now, studies of nonlinear QCD effects on these reactions were done without
taking into account gluon number fluctuations (see \cite{TKM02,GKCN11} and references therein). Such investigation is the aim of the present work.

In this paper we  investigate the consequences of the 
inclusion of gluon number fluctuations in $\gamma^{(*)}\gamma^{(*)}$ collisions. Photon--photon interactions can be understood as a dilute--dilute scattering and thus favours one to rely on the pomeron loop equations. Within the dipole
picture, we demonstrate that in the kinematical range of the  LEP data on total $\gamma\gamma$, $\gamma^*\gamma^*$ cross sections and the real photon structure function $F_2^\gamma(x,Q^2)$, the gluon fluctuations effects are small. However, they contribute significantly in the range which will be probed in the future linear colliders. The paper is organized as follows.  In Section
\ref{sec:amplitudes} we review some important properties of nonlinear high
energy QCD evolution of the dipole-hadron scattering amplitude, in order to
explain how fluctuation effects can be included in this analysis. In Section
\ref{sec:models} we describe the dipole representation of $\gamma\gamma$
scattering and present a model for the dipole-dipole cross section,
the main input of the calculation of observables. Section \ref{sec:results}
is devoted to the description of the available LEP data on two photon collisions
as well as predictions for future experiments. The conclusions are given in
Section \ref{sec:conc}.

\section{QCD dynamics at high energies}\label{sec:amplitudes}

Let us consider the general problem of a scattering between a small dipole
(a colorless quark-antiquark pair) and a dense hadron target, at a given
rapidity interval $Y$. The dipole has transverse size given by the vector
$\rr=\xx-\yy$, where $\xx$ and $\yy$ are the transverse vectors for the quark
and antiquark, respectively, and impact parameter $\bb=(\xx+\yy)/2$. An
important quantity in the description of this process in the high energy
regime is (the imaginary part of) the forward scattering amplitude
$\avg{T(\rr,\bb)}_Y\equiv \avg{T(\xx,\yy)}_Y\equiv \avg{T_{\xx\yy}}_Y$, whose
evolution with rapidity is given by
\begin{equation}\label{eq:b-jimwlk}
\partial_Y \avg{T_{\xx\yy}}_Y = \bar{\alpha}\int d^2z\,
\frac{(\xx-\yy)^2}{(\xx-\zz)^2(\zz-\yy)^2}
\left[\avg{T_{\xx\zz}}_Y+\avg{T_{\zz\yy}}_Y-\avg{T_{\xx\yy}}_Y
-\avg{T_{\xx\zz}T_{\zz\yy}}_Y\right],
\end{equation}
where $\bar{\alpha}=\alpha_sN_c/\pi$ and $\avg{\cdots}$ denotes the average over all the configurations of the target at a given rapidity interval $Y$. This equation is the first equation of a infinite hierarchy, the Balitsky-JIMWLK hierarchy \cite{GIJV10}, and has a simple interpretation in terms of the projectile dipole evolution: if the rapidity is increased by an ammount $\delta Y$, there is a probability for a gluon, with transverse coordinate $\bm{z}$, to be emitted by the quark (or antiquark) of the pair. In the large $N_c$ limit ($N_c$ is the number of colors), this gluon { can be replaced by a quark-antiquark pair at point $\bm{z}$}. This is the dipole picture introduced by Mueller \cite{dipolepic}. Thus, after one step in the evolution, the incoming dipole $(\xx, \yy)$ splits into two new dipoles $(\xx, \zz)$ and $(\zz, \yy)$, which then interact with the target. The last term, $\avg{T(\xx,\zz)T(\zz,\yy)}$, corresponds to the scattering of both new dipoles with the target.

If one performs a mean field approximation,
$\avg{T(\xx,\zz)T(\zz,\yy)}_Y\approx \avg{T(\xx,\zz)}\avg{T(\zz,\yy)}_Y$ and the resulting equation is the BK (Balitsky-Kovchegov) equation \cite{Balitsky,K99a,K99b}, a closed equation for the average dipole scattering amplitude
$\avg{T(\xx,\zz)}_Y\equiv\calN_Y(\rr)$, which, at fixed coupling is given by
 
\begin{equation}\label{eq:bklo}
\partial_Y \calN_Y(\xx,\yy) = \bar{\alpha}\int d^2z\,
\frac{(\xx-\yy)^2}{(\xx-\zz)^2(\zz-\yy)^2}
\left[\calN_Y(\xx,\zz)+\calN_Y(\zz,\yy)-\calN_Y(\xx,\yy)
-\calN_Y(\xx,\zz)\calN_Y(\zz,\yy)\right],
\end{equation}
This equation includes unitarity corrections and is free from the problem of diffusion to the infrared (nonperturbative) region (present in the solution of the linear BFKL equation \cite{bfkl}). Its solution has the following properties: {\tt(i)}
for small $r=|\rr|$, $\calN(\rr)$ is small --the color transparency regime-- and is well approximated by the BFKL solution; {\tt(ii)} for large $r$, the
amplitude approaches the unitarity bound $\calN(\rr)=1$, the so called 'black disc' limit, and the transition between these two regimes takes place at $r=1/Q_s(Y)$. $Q_s(Y)$ is an increasing function of rapidity $Y$ and is called
the {\it saturation scale}, defined { in such a way that} $\calN(\rr)={\cal O}(1)$ (usually $1/2$) when $r=1/Q_s(Y)$.

BK equation has been shown \cite{mp} to belong to the same universality class of the Fisher and Kolmogorov-Petrovsky-Piscounov  (FKPP) equation \cite{fkpp}. Therefore, BK equation admits traveling wave solutions: at asymptotic rapidities, the scattering amplitude depends only on the ratio $r^2Q_s^2(Y)$ instead of depending separately on $r$ and $Y$. This scaling property is called geometric scaling and has been observed in the measurements of the proton structure function at HERA \cite{gscaling}. The amplitude is a wavefront which interpolates between 0 and 1 and travels towards smaller values of $r^2$ with speed $\lambda$ -- the saturation exponent -- keeping its shape, and the saturation scale $Q_s(Y)$ gives the front position.

Within the correspondence between reaction-diffusion processes and
the QCD evolution at high energy, it has been  realized that the Balitsky-JIMWLK hierarchy is not complete because they do not take into account the gluon (dipoles) number fluctuations, which are related to discreteness in the evolution, and thus they are completely missed by BK equation. At least at fixed coupling, fluctuations influence dramatically the QCD evolution at high energies, and so the properties of the scattering amplitudes. Their inclusion results in a new hierarchy of evolution equations, the pomeron loop equations \cite{MS04,IMM04,IT04,MSW05,IT05,LL05}. Because of the complexity of the equations of this new
hierarchy, many of their properties have been known from some approximations \cite{IT04}, after which it has been found that the hierarchy can be generated
from a Langevin equation for the event-by-event amplitude. Formally, this is the BK equation with a noise term, which lies in the same universality class of the stochastic FKPP equation (sFKPP): each realization of the noise means a single realization of the target in the evolution and leads to an amplitude for a single event. Different realizations of the target lead to a dispersion of the
solutions, and then in the saturation momentum $\rho_s\equiv \ln(Q_s^2/k_0^2)$
from one event to another. The saturation scale is now a random variable whose average value is given by
\begin{equation}
\langle Q_s^2(Y) \rangle = \exp{[\lambda^*Y]}
\end{equation}
and the dispersion in the position of the individual fronts is given by
\begin{equation}
\sigma^2 = \langle \rho_s^2 \rangle - \langle \rho_s \rangle^2 = D\bar{\alpha}Y.
\end{equation}
where $D$ is the \textit{diffusion coefficient}, a number expected to be of
order one, which determines the rapidity $Y_D = 1/D$ above which gluon number fluctuations become important.

The probability distribution of $\rho_s$ is, to a good approximation, a Gaussian \cite{Marquet:2006xm}
\begin{equation}
P_Y(\rho_s)\simeq \frac{1}{\sqrt{\pi\sigma^2}}\exp\left[-\frac{(\rho_s-\avg{\rho_s})^2}{\sigma^2}\right].
\end{equation}
For each single event, the evolved amplitude shows a traveling-wave
pattern, which means that geometric scaling is preserved for each
realization of the noise. However, the speed $\lambda^*$ of the wave is smaller
than the speed predicted by BK equation. The average (or physical) amplitude is determined by ($\rho\equiv \ln(1/r^2Q_0^2)$)
\begin{equation}
\avg{\calN(\rho,\rho_s)} = \int^{+\infty}_{-\infty}d\rho_s\,P_Y(\rho_s)\calN(\rho,\rho_s)\,\,,
\end{equation}
with $\calN(\rho,\rho_s)$  being now the event-by-event scattering amplitude.
A crucial property of the physical amplitudes is that at sufficiently high energies, unlike the individual fronts, they will generally not show geometric scaling. More specifically, they  will show additional dependencies upon $Y$, through the front dispersion $\sigma$. Then, geometric scaling is washed out and replaced by the so-called \textit{diffusive scaling} \cite{IT04,IM032,MS04,Hat06}
\begin{equation}\label{eq:avg-amplitude}
\avg{\calN(\rho,\rho_s)} = N\left(\frac{\rho-\avg{\rho_s}}{\sqrt{\bar{\alpha}DY}}\right).
\end{equation}

\begin{figure}[t]
\begin{center}
\scalebox{0.2}{\includegraphics*{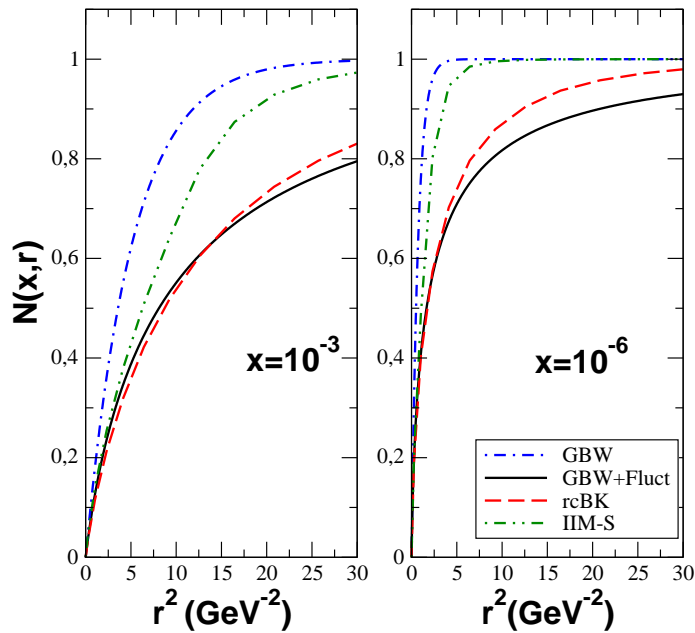}}
\caption{Comparison between the average amplitude $\langle {\cal N}_{GBW} \rangle $ and the event-by-event amplitude predicted by the GBW, IIM-S and rcBK models. The behaviours with $r^2$ are shown
for $x=10^{-3}$ (left panel) and $10^{-6}$ (right panel). }\label{fig:ncomp}
\end{center}
\end{figure}

The different scaling behaviours arising from different versions of QCD evolution were compared with  the available HERA data for inclusive, exclusive and diffractive observables in Ref. \cite{BPRS08} and the quality factor, which estimates the validity of the scaling, was determined. They found that the diffusive scaling leads to best quality factor  for vector meson production at HERA. Furthermore, in \cite{Kozlov07} the authors found that the description of the DIS data is improved once gluon number fluctuations are included and that the values of the saturation exponent and the diffusion coefficient turn out reasonable and agree with values obtained from numerical simulations of toy models which take into account fluctuations.  For instance, for the event-by-event amplitude given by the GBW model \cite{GBW99}
\be\label{eq:gbw}
\calN_{\mbox{GBW}}(r,Y)=1-e^{-r^2Q_s^2(Y)/4},
\ee
where the saturation scale is given by $Q_s^2(Y\equiv\ln(x_0/x))=Q_0^2\left(x_0/x\right)^{\lambda}$,  they have found that $\lambda = 0.225$ and $D = 0.397$ for a $\chi^2/$d.o.f. = 1.14. In contrast, for the $D = 0$ case (no fluctuations), $\lambda = 0.225$  and $\chi^2/$d.o.f. = 1.74. A similar conclusion was obtained considering the IIM model \cite{IIM03} for the event-by-event amplitude, with the values of $\lambda$ and $D$ being quite model independent. { Fig. \ref{fig:ncomp} shows a comparison between the
behaviours  for the scattering amplitude with $r^2$ for given
values of $x$, with and without the inclusion of fluctuations.
Besides GBW model for the event-by-event scattering amplitude we also present the predictions of the IIM-S \cite{iims} and rcBK \cite{rcbk} models used in Ref. \cite{GKCN11}.
 One can see that, 
when the gluon fluctuations effects are included, the onset of saturation is strongly delayed in comparison to the event-by-event scattering amplitude of the GBW model. The same conclusion is valid when $\langle {\cal N}_{GBW} \rangle $   is compared to the IIM-S prediction. In comparison to the rcBK one, the GBW averaged amplitude has a  similar $r^2$ behaviour for $x = 10^{-3}$. However, they become quite different  at smaller values of $x$.}

Although  a definitive conclusion is not possible from the
{ phenomenological} studies presented in Refs. \cite{BPRS08,Kozlov07,AGBSflu,Xiang09,Xiang10-ddis,F11}, they indicate that the presence of the gluon fluctuation effects cannot be disregarded at HERA. Moreover, the contribution of these effects is expected to increase with the energy, which implies that it can be large at LHC. However, due the complexity of the hadron-hadron collisions, it is not clear if the discrimination of these effects will be feasible. Consequently, the search for alternative processes to constrain the presence and magnitude of the gluon fluctuations effects is justified. This will be done in the following through
two-photon scattering processes.

\section{Photon - Photon collisions in dipole representation}
\label{sec:models}

\begin{figure}[ht!]
\begin{center}
\scalebox{0.5}{\includegraphics*{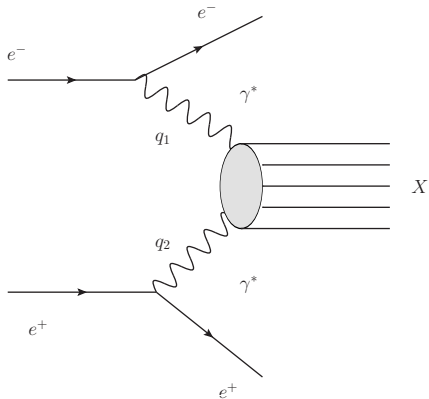}}
\caption{Feynman diagram for the $e^+e^- \rightarrow e^+e^-X$ process.}\label{fig:ee}
\end{center}
\end{figure}

Cross sections of $\gamma^{(*)}\gamma^{(*)}$ scattering can be measured at $e^+e^-$ colliders by tagging both outgoing leptons close to the forward direction (for a review see \cite{Nisius99}).
The process is described by the reaction $e^++e^- \to e^++e^-+X$,
where $X$ is a generic hadronic state formed through the interaction of
photons emitted by the two leptons, and can be represented by the
Feynman diagram in Fig.\ref{fig:ee}, where $q_1$ and $q_2$ are the photon
four-momenta.

\begin{figure}[ht]
\includegraphics*[scale=0.35]{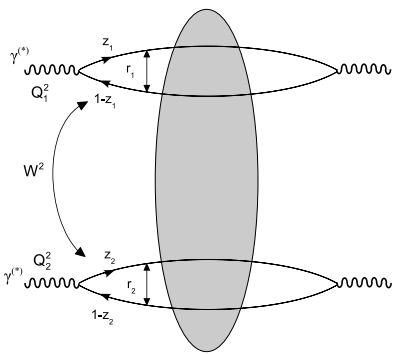}
\caption{Two-photon interactions in the dipole representation.}\label{fig:diag}
\end{figure}

At high energies, the scattering between the two photons can be described in
the dipole frame, in which the photons, with virtualities
$Q_{1,2}^2=-q_{1,2}^2$, fluctuate into quark-antiquark pairs
(two dipoles) with transverse sizes $r_{1,2}$, which then interact and produce
the final state (see Fig.\ref{fig:diag}). Within such formalism, the part of
the two-photon total cross section that determines the energy behaviour  at high energies corresponds to the exchange of gluonic degrees of freedom and is given by \cite{DDR}
\begin{equation}\label{eq:gluon}
\sigma_{ij}
(W^2,Q_1^2,Q_2^2)=\sum_{a,b=1}^{N_f}\int_0^1{\rm d}z_1\int{\rm d}^2
{\bm r}_1|\Psi_i^a(z_1,{\bm r}_1)|^2\int_0^1{\rm d}z_2\int{\rm d}^2
{\bm r}_2|\Psi_j^b(z_2,{\bm r}_2)|^2\sigma_{a,b}^{dd}(r_1,r_2,Y).
\end{equation}
In the above formula, $W^2=(q_1+q_2)^2$ is the collision center of mass squared energy, $z_{1,2}$ are the longitudinal momentum fractions of the quarks in the photons, $\Psi^a_i(z_k,\bm{r})$ denotes the photon wave function, the indices $i,\,j$ label the polarisation states of the virtual photons ($i,\,j=$L or T) and $a,\,b$ label the quark flavours. $\alpha_{em}$ is the electromagnetic coupling constant. The interaction is described by $\sigma_{a,b}^{dd}(r_1,r_2,Y)$, which is the dipole-dipole
cross section. In the eikonal approximation, it can be expressed by
\be\label{eq:ddeikon}
\sigma^{dd}(\rr_1,\rr_2,Y)=2\int{\rm d}^2\bb\,\calN(\rr_1,\rr_2,\bb,Y)
\ee
where $\calN(\rr_1,\rr_2,\bb,Y)$ is the imaginary part of the scattering amplitude for two dipoles with transverse sizes $\rr_1$ and $\rr_2$, relative impact parameter $\bb$ and rapidity separation $Y$. The inclusion of the unitarity corrections in the dipole - dipole scattering was addressed in Ref. \cite{MS96} 
considering independent multiple scatterings between the dipole, with unitarization obtained in a symmetric frame, like the center-of-mass frame. Such  corrections were also estimated considering the Color Glass Condensate formalism in Ref. \cite{IM04}. As in general the applications of the CGC formalism to scattering problems require an asymmetric frame, in which
the projectile has a simple structure and the evolution occurs in the target wavefunction,  the use of the solution of the BK equation in the calculation of the dipole-dipole scattering cross section is not so straightforward.

In order to relate $\calN(\rr_1,\rr_2,\bb,Y)$ with $\calN(\rr,Y)$  we follow  Ref. \cite{GKCN11}, which  considers the model proposed  by Iancu, Kugeratski and Triantafyllopoulos to study the Mueller-Navelet process \cite{IKT08}.
In this model (denoted IKT model hereafter) the dipole-dipole cross section has the following form:
\begin{eqnarray}
\sigma^{dd} (\rr_1,\rr_2,Y) =  2 \pi r_1^2 \calN(r_2,Y_2) \, \Theta(r_1 - r_2) + 2 \pi r_2^2 \calN(r_1,Y_1) \, \Theta(r_2 - r_1) \,\,,
\label{ourmodel}
\end{eqnarray}
where  $Y_i = \ln (1/x_i)$ and 
\begin{eqnarray}
 x_i = \frac{Q_i^2 + 4 m_f^2}{W^2 + Q_i^2}.
\label{xdef}
\end{eqnarray}
The main assumptions in this model are the following: (i) The radial expansion of the gluon distribution in the target (larger dipole) only affects the subleading energy dependence of $\sigma^{dd}$, which implies that it is possible to study the approach towards unitarity limit at a fixed value of the target size; (ii) Only the range $b < R$, where $R =$ Max$(r_1,r_2)$, contributes for the dipole-dipole cross section, {i.e.} it is assumed that ${\cal{N}}$ is negligibly small when the dipoles have no overlap with each other ($b>R$).
As shown in Ref. \cite{GKCN11}, 
due to the quadratic dependence on the size of the larger dipole [See Eq. (\ref{ourmodel})], the contribution of large values of $r_1$ and $r_2$ is quite significant in  the total cross section. It implies that in order to keep our calculations in the perturbative regime a cut in the integration on the pair separation should be assumed. 
As in \cite{GKCN11}, we stop the $r_1$ and $r_2$ integrations at a maximum dipole  size, which 
is chosen to be  $r_{max}  =  {1}/{\Lambda}$, with $\Lambda$ being
a free parameter. As shown in \cite{GKCN11}, this model  successfully describes the current data on the total $\gamma\gamma$ cross section, on the photon structure function $F_2^\gamma(x,Q^2)$ at low $x$ and on the $\gamma^*\gamma^*$ cross section extracted from LEP doubled tagged events, with the expected
value $\Lambda \approx\Lambda_{QCD}$. However, that analysis did not take into account fluctuations effects, which is the aim of this paper.

In what follows we will assume the IKT model for the dipole-dipole cross section and consider that the event-by-event scattering amplitude $\calN(r_i,Y_i)$ is given by Eq. (\ref{eq:gbw}). When taking into account the gluon number fluctuations the scattering amplitude will be replaced by the averaged (physical) amplitude, $\avg{\calN(\rho,\rho_s)}$, which is given by averaging over all possible gluon realizations/events, corresponding to different events in an experiment, Eq. (\ref{eq:avg-amplitude}).

\section{Results}\label{sec:results}

\begin{figure}[t]
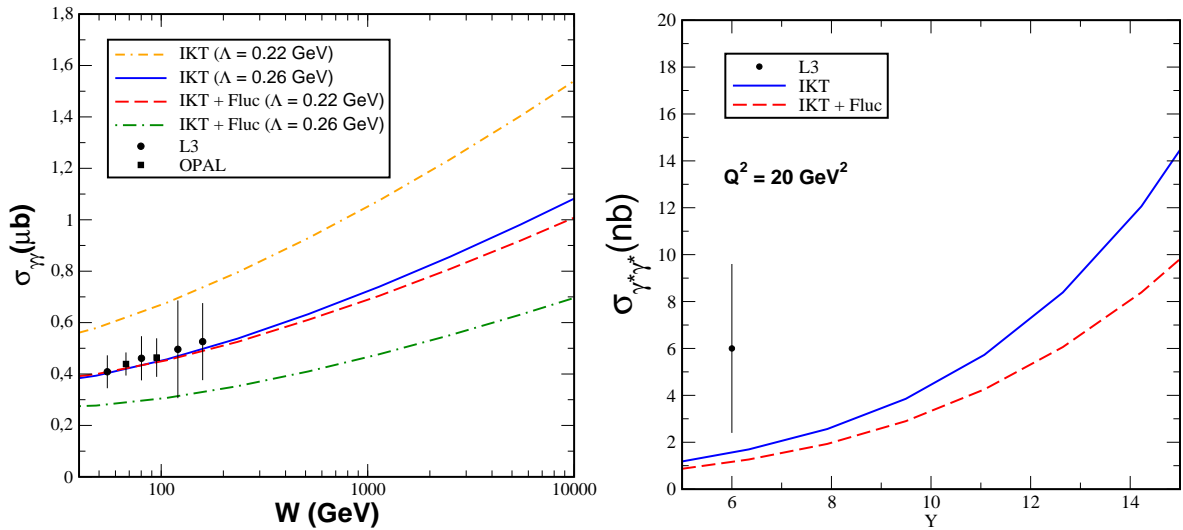

\includegraphics*[scale=0.165]{real_lambdas.eps}
\includegraphics*[scale=0.17]{virtual_ikt.eps}
\caption{(a) Energy dependence of the real $\sigma_{\gamma\gamma}$ cross section. (b) Rapidity dependence of the virtual $\sigma_{\gamma^*\gamma^*}$ cross section at $Q^2_1 = Q^2_2 = Q^2 = 20$ GeV$^2$.}
\label{fig:sigma}
\end{figure}

We use the same values for the parameters $\lambda$, $x_0$ and $D$
obtained in Ref. \cite{Kozlov07} by fitting the $F_2$ HERA data.
Moreover, we assume three flavours with equal masses ($m_f = 0.14$ GeV). The free parameter in our calculations is $\Lambda$, which determines the normalization of the cross sections.  
As in \cite{GKCN11}, we choose  $\Lambda$ in such a way that the experimental data of the real $\gamma \gamma$ cross section \cite{Greal01} in the low energy regime ($W < 60$ GeV) are well fitted. In Fig. \ref{fig:sigma}(a) we present our results for the energy dependence of the real cross section considering the IKT model without  and with fluctuations for different values of $\Lambda$. We can see that it is
not possible to describe the data by using the same value for $\Lambda$ in the two different analyses. The values of $\Lambda$ that allows the description of the data are $\Lambda = 0.26$ GeV (solid line) and  $\Lambda = 0.22$ GeV (dashed line), which are near $\Lambda_{QCD}$, in agreement with our expectations. 
This result can be interpreted as an indication that  the IKT model for the dipole-dipole cross section captures the main features  of the interaction. Furthermore, we observe that the inclusion of the gluon fluctuations effects implies a smother energy behaviour, which agree with the theoretical expectation. However, the difference between the predictions is smaller than 3 $\%$ at $W = 1000$ GeV, which implies that $\sigma_{\gamma \gamma}$ should not be the ideal observable to determine the presence of the gluon fluctuations effects. 

In Fig. \ref{fig:sigma}(b) we present our predictions for the two-photon cross section, as a function of the rapidity $Y\equiv \ln (W^2/Q_1 Q_2)$, for the
case $Q_1^2 = Q_2^2$ (with large $Q_{1,2}^2$) corresponding to the  interaction  of two (highly) virtual photons. In this case, the contribution of the saturation effects is expected to be smaller and, consequently, gluon fluctuation effects to be larger. One can see that gluon number fluctuations diminish the cross section by almost $30 \%$ at $Y = 12$, and their effects
keep increasing at larger values of $Y$. At smaller values of $Q^2$, we verify that both models describe the experimental data with a smaller difference between them. The experimental point is taken from the L3 Collaboration \cite{Gvirt99}.

\begin{figure}[t]
\includegraphics*[scale=0.2]{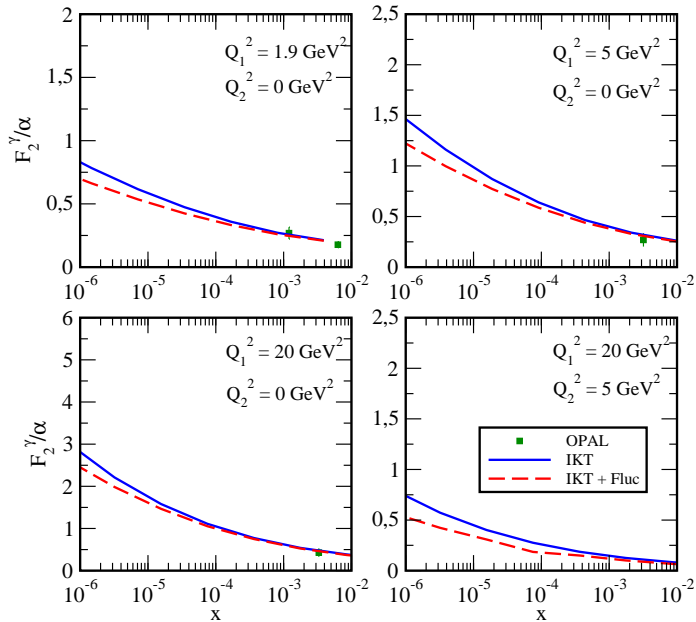}
\caption{The photon structure function $F_2^{\gamma}(x,Q^2)$ as a 
function of  $x=Q^2/(W^2+Q^2)$ for different choices of the photon virtualities.}
\label{fig:f2gama}
\end{figure}

{ Fig. \ref{fig:f2gama} shows} our predictions for the
$x\equiv Q^2/(Q^2+W^2)$ dependence of the photon structure function $F_2^{\gamma}(x,Q^2)$ for different values of the  photon virtualities. The basic idea is that the quasi-real photon structure may be probed by other  photon with a large momentum transfer. We present in the lower right  panel our predictions for the virtual photon structure function. Although there exist only very few data on this observable, its experimental study is feasible in  future linear colliders. The current experimental data \cite{F2g99,F2g00} are described quite well by both models, with the difference between them increasing at small-$x$. In particular,  the difference can be of the order of $30 \%$ in the kinematical range which could be probed in the future linear colliders.

\begin{figure}[ht!]
\includegraphics*[scale=0.15]{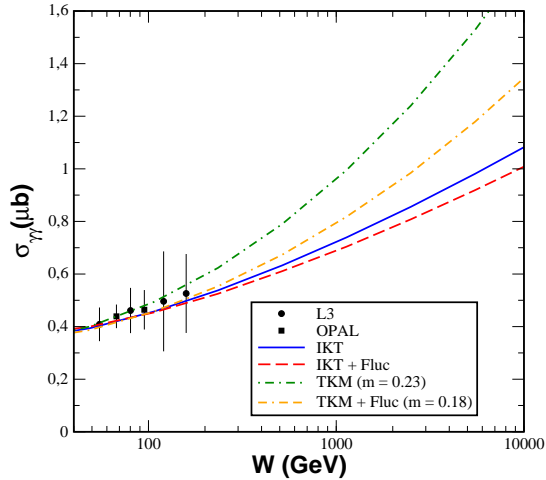}
\caption{Comparison between IKT and TKM models in the description
of the $\sigma_{\gamma\gamma}$ data.}
\label{fig:real_comp}
\end{figure}

In the above analysis we have used IKT model for the dipole-dipole cross section, but the description of this quantity is still an open question. Timneanu, Kwiecinski and Motyka \cite{TKM02}  have proposed a different model (TKM model) to describe $\sigma^{dd}$ in which $\sigma^{dd}(r_1,r_2,Y) = \sigma_0^{dd} \,{{N}}(\rr_{eff}^2,Y)$.  Here $\sigma_0^{dd} = (2/3) \sigma_0$, with  $\sigma_0$ fixed in the GBW model by fitting the $ep$ HERA data and 
$r^2_{eff}=r_1^2r_2^2/(r_1^2+r_2^2)$ being an effective dipole size. The light quark mass $m$ in the photon wave functions is assumed to be a free parameter, to be fixed in such a way to describe the experimental data for $\sigma^{\gamma \gamma}$ at small values of the center-of-mass energy.  Although largely used in the literature, TKM model seems not to be well justified, for it assumes the impact-parameter factorization of the dipole-dipole cross-section, which implies $\sigma^{dd} \propto \sigma_0$. In the dipole-proton case, $\sigma_0$ reflects the size of the proton. However, in the dipole-dipole case, it should reflect the size of the larger dipole. Therefore, taking $\sigma_0^{dd}$ as a constant is an unphysical procedure.  Thus, we believe that  the IKT model is more realistic to describe observables in two-photon interactions. However, this subject deserves more detailed studies. For completeness of the present study, we present in Fig. \ref{fig:real_comp} a comparison between the predictions without and with fluctuations obtained using the TKM model for $\sigma^{dd}$  with the experimental  for the $\gamma\gamma$ cross section. As already observed before, the inclusion of the fluctuation effects implies a smother behaviour with energy. It is important to emphasize that the impact of the fluctuations is larger than observed in the IKT model. We have checked that this conclusion also is valid for the other observables.


\section{Conclusions}\label{sec:conc}

A current open question in the QCD dynamics at high energies 
is if the gluon fluctuations effects should be considered in 
the description of the observables. Results obtained using toy models indicate that these effects are suppressed by the running coupling corrections to the evolution but, since the investigation in QCD of the consequences of these
effects simultaneously is still prohibitive, phenomenological studies which test the implications of the gluon fluctuation effects on observables of different
processes remain important. Following previous studies that indicate that the gluon fluctuation effects may be present in $ep$ collisions at HERA, in this paper we investigated their influence on some of the observables which could be measured in the future linear $e^+ e^-$ colliders. In particular,
we studied the consequences of fluctuations on $\gamma^{(*)}\gamma^{(*)}$
interactions, in the fixed coupling case, within the dipole picture, using
a dipole model for the dipole-dipole cross section. Our results indicate that these effects diminish the increasing with the energy of the real and virtual cross sections and modify the $x$-dependence of the photon structure function. For the total virtual cross section and the photon structure function, the reduction with respect to the case without fluctuations can be of the order of $30 \%$, which implies that  
these effects should not be disregarded in the description of these observables in future colliders. However, because there are large uncertainties present in the current analyses---in particular, the normalization of cross sections, associated with the choices of quark masses and the parameter $\Lambda$, and even in the models for the dipole-dipole cross section $\sigma_{dd}$---observing the presence of the fluctuation effects on inclusive processes in photon-photon collisions will be a hard task.

\section*{Acknowledgments}

This work is partially supported by CNPq and FAPERGS.

\bibliographystyle{unsrt}
\bibliography{gg_refs}

\end{document}